

\magnification\magstep1
\parskip=\medskipamount
\hsize=6 truein
\vsize=8.2 truein
\hoffset=.2 truein
\voffset=0.4truein
\baselineskip=14pt
\tolerance=500

\font\titlefont=cmbx12
 at 10 truept
\font\authorfont=cmcsc10
\font\addressfont=cmsl10 at 10 truept
\font\smallbf=cmbx10 at 10 truept
 4
\def\shalf{\hbox{${\textstyle{1\over 2}}$}}

\outer\def\beginsection#1\par{\vskip0pt plus.2\vsize\penalty-150
\vskip0pt plus-.2\vsize\vskip1.2truecm\vskip\parskip
\message{#1}\leftline{\bf#1}\nobreak\smallskip\noindent}

\newcount\notenumber

\def\note{\advance\notenumber by 1
\footnote{$^{\the \notenumber}$}}

\newdimen\itemindent \itemindent=13pt
\def\textindent#1{\parindent=\itemindent\let\par=\resetpar%
\indent\llap{#1\enspace}\ignorespaces}

\let\oldpar=\par
\def\resetpar{\oldpar\parindent=0pt\let\par=\oldpar}

\font\ninerm=cmr9 \font\ninesy=cmsy9
\font\eightrm=cmr8 \font\sixrm=cmr6
\font\eighti=cmmi8 \font\sixi=cmmi6
\font\eightsy=cmsy8 \font\sixsy=cmsy6
\font\eightbf=cmbx8 \font\sixbf=cmbx6
\font\eightit=cmti8
\def\eightpoint{\def\rm{\fam0\eightrm}
  \textfont0=\eightrm \scriptfont0=\sixrm \scriptscriptfont0=\fiverm
  \textfont1=\eighti  \scriptfont1=\sixi  \scriptscriptfont1=\fivei
  \textfont2=\eightsy \scriptfont2=\sixsy \scriptscriptfont2=\fivesy
  \textfont3=\tenex   \scriptfont3=\tenex \scriptscriptfont3=\tenex
  \textfont\itfam=\eightit  \def\it{\fam\itfam\eightit}%
  \textfont\bffam=\eightbf  \scriptfont\bffam=\sixbf
  \scriptscriptfont\bffam=\fivebf  \def\bf{\fam\bffam\eightbf}%
  \normalbaselineskip=9pt
  \setbox\strutbox=\hbox{\vrule height7pt depth2pt width0pt}%
  \let\big=\eightbig \normalbaselines\rm}
\catcode`@=11 %
\def\eightbig#1{{\hbox{$\textfont0=\ninerm\textfont2=\ninesy
  \left#1\vbox to6.5pt{}\right.\n@space$}}}
\def\vfootnote#1{\insert\footins\bgroup\eightpoint
  \interlinepenalty=\interfootnotelinepenalty
  \splittopskip=\ht\strutbox %
  \splitmaxdepth=\dp\strutbox %
  \leftskip=0pt \rightskip=0pt \spaceskip=0pt \xspaceskip=0pt
  \textindent{#1}\footstrut\futurelet\next\fo@t}
\catcode`@=12 %

\def\v{\vec}
\def\R{{\cal R}}
\def\h#1{\hat{#1}}


\rightline{Freiburg, THEP-96/7}
\rightline{gr-qc/9605011}
\bigskip
{\baselineskip=24 truept
\titlefont
\centerline{CONSISTENTLY IMPLEMENTING THE FIELDS SELF-ENERGY}
\centerline{IN NEWTONIAN GRAVITY}
}

\vskip 1.1 truecm plus .3 truecm minus .2 truecm

\centerline{\authorfont Domenico Giulini\footnote*{
e-mail: giulini@sun2.ruf.uni-freiburg.de}}
\vskip 2 truemm
{\baselineskip=12truept
\addressfont
\centerline{Fakult"t f\"ur Physik,
Universit\"at Freiburg}
\centerline{Hermann-Herder Strasse 3, D-79104 Freiburg, Germany}
}
\vskip 1.5 truecm plus .3 truecm minus .2 truecm

\centerline{\smallbf Abstract}
\vskip 1 truemm
{\baselineskip=12truept
\leftskip=3truepc
\rightskip=3truepc
\parindent=0pt

{\eightpoint
We consider in a pedagogical fashion alterations to Newtonian
gravity due to the postulate that all energy corresponds to
active gravitational mass when applied to the self-energy of
the gravitational field. We show why a simple addition of
${1\over c^2}$ times the gravitational field energy to the
matter density in Newton's field equation is inconsistent.
A consistent prescription is shown and discussed.
The connection to general relativity is pointed out.
\par}}

\beginsection{1. Introduction}

The issue addressed in this letter arises if one wishes to model
the self-coupling of the gravitational field within Newtonian
gravity. Simple non-linear alterations of Newton's field equation
are often employed as simplified models for general relativity.
The purpose of this letter is to show how this can be done and to 
point out certain flaws in the usually accepted prescription, 
as for example given in [1][2].

To be more precise, we recall that the Newtonian gravitational field, 
$\varphi$, and the density of (ponderable) matter, $\rho$, obey
$$
\Delta\varphi=4\pi G\rho
\eqno{(1.1)}
$$
where $G$ is Newton's gravitational constant. The force per unit volume
is given by
$$
\vec f=-\rho\vec\nabla\varphi .
\eqno{(1.2)}
$$
Together these equations imply that in order to build up a field
$\varphi$ from $\varphi=0$ one has to invest the work
$$
A=-{1\over 8\pi G}\int_{R^3} \Vert\vec\nabla\varphi\Vert^2\,dV.
\eqno{(1.3)}
$$

If we add the assumption that all energy acts as active gravitational
mass, according to $E=mc^2$, and also think of the integrand in
(1.3) as representing energy density, we might be tempted to
consider the modified equation
$$
\Delta\varphi=
4\pi G\left(\rho-{1\over 8\pi G c^2}\Vert\vec\nabla\varphi
                                      \Vert^2\right)
\eqno{(1.4)}
$$
with the aim to incorporate into Newtonian gravity the
following
\proclaim Principle (P). All energy acts as
source for the gravitational field.\par
A field equation satisfying {\bf P} must be non-linear.
One might wonder whether (1.4) gives a Newtonian model that
satisfies {\bf P}. If it were true that it shared this
qualitative feature with general relativity one might
profitably employ this single scalar equation to study certain
qualitative features of general relativity in a mathematically simpler
environment. In fact, (1.4) is often proposed in pedagogical
discussions to precisely this end. For example, in [2][3] the
authors suggest that some useful lessons concerning the
energy-regulating power of the gravitational field can be
learned from model theories of charged particles based on
(1.4). In passing we remark that (1.4) can be written in
a linear form by making the field-redefinition
$\psi:=\exp(\varphi/2c^2)$:
$$
\Delta\psi={2\pi G\over c^2}\rho\psi 
\eqno{(1.5)}
$$
where the boundary conditions $\varphi(r\rightarrow\infty)=0$
translate to $\psi(r\rightarrow\infty)=1$.
In the following we shall for simplicity always assume $\rho$ to
have compact support $B\subset R^3$.

In section 3 we discuss what is wrong with a theory based on
(1.2,4) and suggest a different and consistent theory in section~4.
Section~5 briefly discusses some properties of spherically
symmetric solutions to the latter and section~6
points out the relation to general relativity. Section~2
summarizes some facts from Newtonian gravity. We employ the
standard summation convention for repeated indices in up-down
positions and use the euclidean metric $\delta_{ab}$
to raise and lower indices. $\nabla_a$ denotes the partial
derivative with respect to $x^a$. 3-component vectors are also
written with an arrow, $\vec \xi$, with
$\vec\xi\cdot\vec\eta=\xi_a\eta^a$ denoting the scalar product.

\beginsection{2. Newtonian Recollections}

To see what is wrong with (1.4) it is helpful to first give a derivation
of (1.3). Consider a one-parameter family of diffeomorphisms 
$s\mapsto\sigma_s$ such that $\sigma_{s=0}=id$ and
${d\over ds}\vert_{s=0}\sigma_s(\vec x)=\vec\xi(\vec x)$.
We wish to use $\sigma_s$ to redistribute the matter by dragging 
it along this flow. Pulling back the 3-form $\rho dV$ by the 
inverse diffeomorphisms we obtain
$\rho_sdV:= (\sigma_{s}^{-1})^*(\rho dV)$ and hence for the
Lie-derivative of the density $\rho$ along $\vec\xi$
$$
\delta\rho:=
{d\rho_s\over ds}\Big\vert_{s=0}=-\vec\nabla\cdot(\rho \vec\xi)
\eqno{(2.1)}
$$
where here and in the following we use the variational symbol,
$\delta$, for the derivative at $s=0$ and call it `the variation'
of the quantity in question.

The variation of the work done to the system is easily
determined using (1.2):
$$
\delta A = -\int_{R^3}\vec \xi\cdot \vec f\,dV
         = -\int_B \varphi\vec\nabla\cdot(\rho\vec\xi)\,dV.
\eqno{(2.2)}
$$
Equations (2.1,2) imply
$$
\delta A=\int_B\varphi \delta\rho\, dV.
\eqno{(2.3)}
$$
This equation is independent of the field equation. If we assume
the validity of (1.1) throughout the (adiabatic) motion we can use
it to eliminate $\delta\rho$ and write (2.3) solely in terms of
$\varphi$. The result (1.3) then easily follows.

 From (1.1,2) it follows that the force per unit mass may be
derived from a symmetric stress tensor, $f_a=-\nabla^b t_{ab}$,
where
$$
t_{ab} ={1\over 4\pi G}\left((\nabla_a\varphi)(\nabla_b\varphi)
        -\shalf\delta_{ab}\Vert\v\nabla\varphi\Vert^2\right)
\eqno{(2.4)}
$$
so that
$$
\delta A = -\int_{R^3} f_a\xi^a\, dV
         =  \int_{R^3}\xi^a\nabla^bt_{ab}\, dV
         =  \int_{R^3}\nabla^{(a}\xi^{b)}t_{ab}\, dV.
\eqno{(2.5)}
$$
Here we assumed $\Vert\vec\xi(r\rightarrow\infty)\Vert<ar$ for some
real constant $a$ and that derivatives of $\varphi$ fall off as fast 
as $r^{-2}$. Vector fields which satisfy $\nabla^{(a}\xi^{b)}=0$
(Killing equation) generate rigid motions and are given by  
$\vec\xi(\vec x)=\vec k$ (translations) and 
$\vec\xi(\vec x)=\vec k\times\vec x$ (rotations), for constant $\vec k$.
For those $\delta A=0$, as it must be by
the principle of action~=~reaction. Otherwise the system would 
self-accelerate.

Finally we define the gravitational mass as the total flux of
the gravitational field $\vec\varphi$ out to infinity:
$$
M_g:=\mathop{\lim}_{r\rightarrow\infty}
     {1\over 4\pi G}\int_{S^2_r}\vec n\cdot\vec\nabla\varphi\,do.
\eqno{(2.6)}
$$
$S^2_r$ denotes a two-sphere of radius $r$, $\vec n$ its 
(outward pointing) normal, and $do$ the surface element on $S^2_r$. 
The limit of integrals in (1.14) is sometimes abbreviated
by  $\int_{S^2_{\infty}}$.

\beginsection{Why Inconsistent?}

Since in Newtonian theory $M_g=M_m:=\int \rho dV$, $M_g$ only
depends on the amount but not on the distribution of matter and
clearly {\bf P} cannot be satisfied. Now, replacing (1.1) by (1.4),
one obtains the following formula for the variation $\delta M_g$
$$
\delta M_g = \int_{B}\sum_{n=0}^{N-1}{1\over n!}
             \left({\varphi\over c^2}\right)^n\,\delta\rho\,dV
           + {1\over N! c^{2N}}{1\over 4\pi G}\int_{R^3}\varphi^N\,
             \delta(\Delta\varphi)\,dV
\eqno{(3.1)}
$$
where we have used (1.4) $N$ times to replace $\Delta\varphi$.
For a regular matter distribution $\varphi$ will be bounded,
say $\varphi(\vec x)< K\,,\forall\vec x\in R^3$. Also, the integral 
over ${1\over 4\pi G}\delta(\Delta\varphi)$ just represents 
$\delta M_g$ so that the last term on the right hand side is 
majorized by ${1\over N!}(K/c^2)^N\,\delta M_g$. It vanishes in
the limit $N\rightarrow \infty$. In this limit the sum on the right
side is just the exponential function. Thus we obtain the result:
$$
\delta M_g = \int_B\delta\rho\,\exp(\varphi/c^2)\,dV
\eqno{(3.2)}
$$
which, recalling (2.3), deviates from $\delta A/c^2$ by all the
higher-than-linear terms in the expansion of the exponential.
Hence (1.2,4) violates {\bf P}. This is not really surprising, since
(1.3) was derived under the assumption of (1.1,2). Changing it
to (1.2,4) also invalidates (1.3). A correct procedure
must iterate the step that led from (1.1) to (1.4). For example,
the next (second) step would be to determine a modified
expression for the gravitational field energy from (1.2,4)
and then change (1.4) accordingly. Eventually this procedure
should converge to a self-consistent field equation.
However, as we will see in the next section, such a
self-consistent field equation can actually be guessed
directly.

At the end of this section we also point out another flaw in the
combination (1.2,4). Using (1.5) to replace $\rho$ in (1.2) one
easily derives
$$
f_a=-\exp(-\varphi/c^2)\nabla^b(\exp(\varphi/c^2)\,t_{ab})
\eqno{(3.3)}
$$
with $t_{ab}$ from (2.4). From this expression it follows that
the force density is not the divergence of a stress tensor.
There are many ways to isolate the part that obstructs the
right hand side of (2.11) to be written as the divergence
of a symmetric tensor. Two obvious ways are
$$\eqalignno{
f_a =& -\nabla^b t_{ab} - {1\over 8\pi Gc^2}
        \Vert\vec\nabla\varphi\Vert^2\nabla_a\varphi
&(3.4a)\cr
    =& - \nabla^b\left[(1+\varphi/c^2)t_{ab}\right]
       + {1\over 8\pi G c^2}\Delta\varphi\nabla_a\varphi^2.
&(3.4b)\cr}
$$
The system (1.2,4) thus potentially violates the principle
action~=~reaction.\note{To manifestly show a violation one should
prove existence of a regular solution to (1.4) with
$\varphi(r\rightarrow\infty)=0$ for which $\int f_a\xi^a\not =0$
for some generator $\vec\xi$ of a rigid motion.}

\beginsection{4. A Consistent Modification}

Equation (3.2) was derived assuming (1.4) but not (1.2). If we
maintain (1.4,5) but call $\phi=c^2\exp(\varphi/c^2)$ rather
than $\varphi$ the gravitational potential, we have (3.2)
just expressing the validity of {\bf P}, i.e. $c^2\delta M_g=\delta A$
with $\delta A$ given by (2.3). This re-interpretation
implies that (1.2) has to be replaced by
$$
\vec f=-\rho\vec\nabla\phi 
\eqno{(4.1)}
$$
and that (1.4,5) written in terms of $\varphi$ reads
$$
\Delta\phi 
= {4\pi G\over c^2}\left(\rho\phi
+ {c^2\over 8\pi G}{\Vert\vec\nabla\phi\Vert^2\over\phi}\right).
\eqno{(4.2)}
$$
To be sure, for explicit calculations one would preferably use
(1.5) where $\psi=c\sqrt{\phi}$. $\phi$ must satisfy the
boundary conditions $\phi(r\rightarrow\infty)=c^2$. The
Newtonian approximation is obtained from expanding
$\phi=c^2+\varphi+O(\varphi^2)$ and keeping only linear terms in
(4.2). Note that in the expression (2.6) for $M_g$ we must
write $\phi$ instead of $\varphi$. But for $r\rightarrow\infty$
only the linear term in $\varphi$ contributes to the surface
integral so that (3.2) is still valid. This is why (3.2) indeed
expresses the validity of {\bf P} for (4.1,2). To be sure, once (4.1,2)
are established, the equation $c^2\delta M_g=\delta A$ is most easily
proven directly. For completeness we give a short direct proof in the
appendix. The point of our derivation of (3.2) was that it suggested
the definition of $\phi$ in terms of $\varphi$ and hence (4.2).
It is interesting to note that (4.2) is precisely the equation
that Einstein already proposed before the advent of general
relativity in 1912 [4].

Equations (4.1,2) also manifestly implie the principle
action~=~reaction in the sense above. Indeed, we now have
$f_a=-\rho\exp(\varphi/c^2)\nabla_a\varphi$. Replacing
$-\rho\nabla_a\varphi$ by the right hand side of (3.3) just 
cancels the exponential function outside the derivative so that 
the remaining divergence can be rewritten in terms of $\phi$.
This leads to the desired formula, $f_a=-\nabla^b t_{ab}$, with
$$
t_{ab} = {1\over 4\pi G c^2}\left\{{1\over\phi}\left[
            (\nabla_a\phi)(\nabla_b\phi)-\shalf\delta_{ab}
            \Vert\v\nabla\phi\Vert^2\right]\right\}.
\eqno{(4.3)}
$$

We may interpret the two terms on the right hand side of (4.2) as
energy densities due to ponderable matter and the gravitational 
field respectively. The sum of both determines the convergence
$\Delta\phi$ of the gravitational field $-\vec\nabla\phi$. Both
terms are positive since $\phi$ is positive. This is in contrast to
(1.4), where the Newtonian gravitational field energy was negative
definite, which is usually said to have its origin in the attractivity
of gravity. But of course here gravity is also attractive. What is
different here is that the (rest-) energy of matter depends on the
value of the gravitational potential at its location. This allows that
a contraction of a matter distribution enhances the field energy although
the total energy decreases.  This is achieved by displacing the matter
into regions of smaller gravitational potential and thereby sufficiently
decreasing the matter part of the energy.

The total gravitational energy is given by
$$
E_{\rm total} := c^2M_g = \int_{B}\rho\phi\, dV
       + {c^2\over 8\pi G}\int_{R^3}
         {\Vert\vec\nabla\phi\Vert^2\over\phi}\,dV
       =:  E_{\rm matter}+E_{\rm field}
\eqno{(4.4)}
$$
where the expression for $E_{\rm field}$ can also be written in terms
of an integral over $B$ (the support of $\rho$) only. To see this we
recall that for large distances from the source we have the expansions
for $\phi$ and $\psi$
$$\eqalignno{
{\phi\over c^2} &= 1-{GM_g\over c^2r} + O(r^{-2})
     &(4.5a)\cr
\psi &= 1-{GM_g\over 2c^2r} + O(r^{-2})
     &(4.5b)\cr}
$$
so that $E_{\rm total}$ can also be expressed as an integral
of ${c^4\over 2\pi G}\Delta\psi=c^2\rho\psi$ (using (1.5)) over $B$.
Replacing $E_{\rm total}$ by this expression in
$E_{\rm field}=E_{\rm total}-E_{\rm matter}$ one obtains
$$
E_{\rm field} = c^2\int_{B} \rho\sqrt{\phi\over c^2}
      \left(1-\sqrt{\phi\over c^2}\right)\,dV.
\eqno{(4.6)}
$$

\beginsection{5. Solution for Homogeneous Spherical Star}

In this section we determine the gravitational field for the 
externally prescribed mass distribution 
$$
\rho = \cases{{3M_m\over 4\pi R^3} & for $r<R$      \cr
                                   &                \cr
              0                    & for $r\geq R$  \cr}
\eqno{(5.1)}
$$
where $M_m$ is the total (`bare') mass of matter:
$M_m=\int_B\rho\,dV$. It will be convenient to introduce the `matter
radius' $\R_m$ and the `gravitational radius' $\R_g$:
$$
\R_m={GM_m\over c^2}, \quad \R_g={GM_g\over c^2}
\eqno{(5.2)}
$$
and the abbreviation
$$
\omega = \sqrt{3\R_m\over 2R}\,{1\over R}.
\eqno{(5.3)}
$$
We use (1.5), set $\psi(r)=\chi(r)/r$, and obtain
$$
   \chi''=\cases{\omega^2\,\chi & for $r<R$      \cr
                     0          & for $r\geq R$. \cr}
\eqno{(5.4)}
$$
The general solution which makes $\phi$ (and hence $\psi$)
finite at $r=0$ is
$$
\psi(r)=\cases{K\,{\sinh(\omega r)\over r} 
& for $r<R$     \cr 
1-{{\R}_g\over 2r}      
& for $r\geq R$. \cr}
\eqno{(5.5)}
$$
The integration constants $K$ and $\R_g$ are determined
by the requirement that $\phi$ (and hence $\psi$) should be
continuously differentiable at $r=R$:
$$\eqalignno{
\R_g &= 2R\left[1-{\tanh(\omega R)\over\omega R}\right]
&(5.6)\cr
K&={1\over\omega\cosh(\omega R)}.
&(5.7)\cr}
$$

Fixing the radius $R$ in $(4.6)$ gives us $\R_g$ as function of $\R_m$, 
i.e., the gravitational mass as function of the bare mass.
In terms of the dimensionless quantities $y=\R_g/R$ and $x=\R_m/R$
this reads:
$$
y = f(x) = 
2\left[1-{\tanh(3x/2)^{1\over 2}\over(3x/2)^{1\over 2}}\right]
\eqno{(5.8)}
$$
which for $x\geq 0$ maps monotonically $[0,\infty]\rightarrow[0,2]$.
For small $x$ one has $f(x)=x-{3\over 5}x^2+{51\over 140}x^3+O(x^4)$.
The fact that $f(x)<2\quad \forall x\in R_+$ means that the 
gravitational mass is bounded by a quantity depending only on the 
geometry (here $R$) of the mass distribution:
$$
M_g<R{2c^2\over G}.
\eqno{(5.9)}
$$
Note that this is achieved with all contributions to the 
gravitational mass on the right hand side of $(3.3)$ being
positive. No subtractions are taking place. Rather, high matter 
densities $\rho$ are suppressed by the small potentials $\phi$
produced by them (i.e. `red-shifted' in general relativistic
terminology). This can be seen in detail from the following
expressions:
$$\eqalignno{
E_{\rm total}
& = {2Rc^4\over G}\left[1-{\tanh(\omega R)\over\omega R}\right]
&(5.10a)\cr
& = M_mc^2\left[1-{3\R_m\over 5R}+O(\R_m^2/R^2)\right]
&(5.10b)\cr
E_{\rm matter} 
& = {Rc^4\over G}\left[{\tanh(\omega R)\over\omega R}
     +\tanh^2(\omega R)-1\right]
&(5.11a)\cr
& = M_mc^2\left[1-{6\R_m\over 5R}+O(\R_m^2/R^2)\right]
&(5.11b)\cr
E_{\rm field}
& = {3Rc^4\over G}\left[1-{\tanh(\omega R)\over\omega R}
     -{1\over 3}\tanh^2(\omega R)\right]
&(5.12a)\cr
& = M_mc^2\left[{3\R_m\over 5R}+O(\R_m^2/R^2)\right]
&(5.12b)\cr}
$$
where the second expressions on the right hand sides are
expansions of the first in terms of $\R_m/R$.
Also, recall that $\R_mc^4/G=M_mc^2$. Note the familiar
${5\over 3}$-term in (5.10b) for the Newtonian binding
energy.

Decreasing $R$ for fixed $\R_m$ we see from (5.10b-12b) that
to first approximation this enhances the field energy and at
the same time decreases the matter energy twice as fast, so as to
decrease the total energy by the same amount by which
the field energy increased. Clearly the total energy must decrease in 
accordance with the attractivity of the gravitational interaction.

Coming back to (5.9) we next show that it remains valid for any
spherically symmetric matter distribution. In particular, it
remains valid for more realistic matter distributions (of compact
support $r<R$) which are determined by a coupled system of (4.2)
with some equations of state for the matter. The proof is simply
this: For $r\geq R$ (5.4) is solved by $\chi_{+}(r)=r-\R_g/2$ and
by some function $\chi_{-}(r)$ for $r\leq R$. Continuity and
differentiability of $\phi$ at $r=R$ is equivalent to
$$\eqalignno{
\chi_{-}(R)  &= R-\shalf\R_g  &(5.13)\cr
\chi'_{-}(R) &= 1.            &(5.14)\cr}
$$
Suppose $\chi(R)\leq 0$, then $\chi''={2\pi G\over c^2}\rho\chi$
with $\rho\geq 0$ implies $\chi''(r)\leq 0$ for all $r\leq R$,
with strict inequality if $r$ lies in the support
of $\rho$. Equations $(5.13,14)$ now imply that the curve
$r\rightarrow\chi_-(r)$ lies below the curve $r\rightarrow r-\shalf\R_g$
for $r\leq R$, which in turn implies $\chi(r=0) <-\shalf\R_g<0$, where
the last inequality just expresses the positivity of the gravitational
mass. But this contradicts the regularity of the gravitational potential
which requires a finite value of $\psi(r=0)$ and thus $\chi(r=0)=0$.
Hence we must have $\chi(R)>0$ or, by $(5.13)$, $\R_g<2R$.

Finally we mention that the finite (negative) bare mass for the
point-charge model of ref. [V] crucially depends on taking $\varphi$
(and (1.4)) rather than $\phi$ as gravitational potential. In the
latter case it unfortunately turns out to be infinite. At this
state of affairs the finite bare mass obtained in [V] is not a
consequence of {\bf P}, but rather an artifact of (1.4), which
violates {\bf P}. But {\bf P} seems crucial for any model of
general relativity and hence (1.4) does not seem suited to model
general relativity for questions concerning the energy regulating
power of the gravitational field.

\beginsection{6. Connection with General Relativity}

Let us consider a Lorentz metric in which only the time-time 
component of the metric differs from its Minkowski value. We write
$$
ds^2=-2\phi\, dt^2+d\vec x\cdot d\vec x
\eqno{(6.1)}
$$
and require that for large spatial distances this approaches the
Minkowski metric:
$$
\lim_{\Vert\vec x\Vert\rightarrow\infty}\phi(\vec x)=\shalf c^2.
\eqno{(6.2)}
$$
Using the parameter $t$, the equations for a timelike
geodesic curve boil down to ($\dot{\vec x}=d\vec x/dt$)
$$
\ddot {\vec x}=-\vec\nabla\phi+\phi^{-1}
(\dot{\vec x}\cdot\vec\nabla\phi)\dot{\vec x}
\eqno{(6.3)}
$$
which, neglecting terms $\propto (v/c)^2$ for the moment, are just the
Newtonian equations of motion for a point mass in the external
potential $\phi$. Setting for the moment $2\phi=\psi^2$, the
components of the Ricci tensor are most easily calculated with
respect to the orthonormal tetrad:
$e_{\h t}=\psi^{-1}\partial_t$, $e_{\h a}=\partial_a$ for $a=1,2,3$
(the hat over the indices signifies the orthonormality). We obtain:
$$\eqalignno{
R_{\h a\h c}
& =
-\psi^{-1}\psi_{,a,c}
&(6.4)\cr
R_{\h t\h t}
& =
\psi^{-1}\Delta\psi
&(6.5)\cr}
$$
where $\Delta$ is just the ordinary Laplacian $\partial^a\partial_a$
and $\partial_a\psi=\psi_{,a}$. The scalar curvature then follows
(summation over $\h a$):
$$
R=R_{\h a\h a}-R_{\h t\h t}=-2\psi^{-1}\Delta\psi.
\eqno{(6.6)}
$$

Let us now consider Einstein's equations
$$
R_{\mu\nu}={8\pi G\over c^4}\left(T_{\mu\nu}
            -\shalf g_{\mu\nu}T^{\lambda}_{\lambda}\right)
\eqno{(6.7)}
$$
with an energy momentum tensor
$T_{\mu\nu}=\rho u_{\mu}u_{\nu}$ for stationary pressureless dust of
rest-mass-density $\rho$ and 4-velocity $u=c\psi^{-1}\partial_t$. 
If we varied the Einstein-Hilbert action only within the class
of metrics of the form (6.1) we would obtain only the
$\h t\h t$-component of (6.7), which explicitly
reads: $\Delta\psi={4\pi G\over c^2}\rho\psi$. In terms of
$\phi$ we have
$$
\Delta\phi 
= {8\pi G\over c^2}\left(\rho\phi
+ {c^2\over 16\pi G}{\Vert\vec\nabla\phi\Vert^2\over\phi}\right).
\eqno{(6.8)}
$$
This is almost (4.3) except for an additional factor of two
in the $G$-dependence. But note that the boundary condition
(6.2) differs from (4.5a) by a factor \shalf which implies
that (6.8) and (4.2) have the same Newtonian limit.
A solution of (4.2), like (5.5), can be easily turned
into a solution to $(5.6)$ if we multiply it by $\shalf$ and
replace $G$ by $2G$.
Thus we conclude that $(3.3)$ is essentially the time-time-component
of Einstein's equations. Note however that we cannot solve the
full set of Einstein's equations with the ansatz $(5.1)$. In fact,
adding the trace of the spatial part
to the $\h t\h t$-part, the left hand side is zero according to
$(5.4,5)$ whereas the right side is easily seen to be proportional 
to $T_{\h t\h t}$. So there is no solution, at least if the matter 
satisfies the dominant energy condition
$\vert T_{\h t\h t}\vert\geq \vert T_{\h a\h c}\vert$. Note that the
trace of Einstein's equations $(5.7)$, $R=-{8\pi G\over c^4}T$, is
already in contradiction to their $\h t\h t$-component, since
$T=T_{\h a\h a}-T_{\h t\h t}= -T_{\h t\h t}=-\rho c^2$ so that
with $(5.6)$ the trace part reads
$\Delta\psi=-{4\pi G\over c^2}\psi$ which differs in sign from the
equation above. This shows how Einstein's scalar equation of 1912
[4] is related to the time-time-component of the general
relativistic tensor equation.

\beginsection{Appendix}

In this appendix we wish to give a short direct proof that
(4.2) satisfies {\bf P}, i.e., that $\delta A = c^2\delta M_g$.

Using the generally valid equation (2.3), now with $\phi$
replacing $\varphi$, we must eliminate $\rho$ via (3.3).
This is most easily done if we set $\phi=c^2\psi^2$ and
use (1.5). We obtain:
$$\eqalignno{
\delta A  =& {c^4\over 2\pi G}\int_{R^3} \psi^2\,
       \delta\left[\Delta\psi\over\psi\right]\,dV
    =  {c^4\over 2\pi G}\int_{R^3}\left[\psi\Delta(\delta\psi)
                      -(\Delta\psi)\,\delta\psi\right]\, dV
&(A.1)\cr
    =& {c^4\over 2\pi G}\int_{S^2_{\infty}}
        \vec n\cdot\left[\psi\vec\nabla(\delta\psi)
              -(\vec\nabla\psi)\,\delta\psi\right]\, do.
&(A.2)\cr}
$$
Now, the conditions for large $r$ imply that
$\vec\nabla\psi$ falls off as fast as $1/r^2$ and $\delta\psi$ as fast
as $1/r$. Hence the second term in the last bracket does not
contribute. Therefore we may reverse its sign and obtain
$$
\delta A = {c^4\over 2\pi G}\ \delta\int_{S^2_{\infty}}
     (\vec n\cdot\vec\nabla\psi)\psi\,dV
  = {c^2\over 4\pi G}\ \delta\int_{S^2_{\infty}}
      \vec n\cdot\vec\nabla\phi\,dV = c^2\delta M_g.
$$

\beginsection{References}

\item{[1]}
Geroch, R. (1978): ``On the positive mass conjecture''.
In: Theoretical Principles in Astrophysics and Relativity,
pp 245-252. Edited by N.R. Lebovitz, W.H. Reid,
and P.O. Vandervort, University of Chicago Press.

\item{[2]}
Visser, M. (1989):
``A Classical Model For The Electron''.
{\it Phys. Lett. A} {\bf 139}, 99-102.

\item{[3]}
Robinson, T.R. (1995):
``Mass and charge distributions of the classical electron''.
{\it Phys. Lett. A} {\bf 200}, 335-339.

\item{[4]}
Einstein, A. (1912):
``Zur Theorie des statischen Gravitationsfeldes''.
{\it Ann. Phys.} (Leipzig) {\bf 38}, 443-458.

\end